# Title:

# GET: A Generative EEG Transformer for continuous context-based neural signals.

# Authors:

Omair Ali[1,4,5], Muhammad Saif-ur-Rehman[3,5], Marita Metzler[1], Tobias Glasmachers[2], Ioannis Iossifidis[3] and Christian Klaes[1]

# Author affiliation:

[1] Faculty of Medicine, Department of Neurosurgery, University hospital Knappschaftskrankenhaus Bochum GmbH, Germany, [2]Institut für Neuroinformatik, Ruhr University Bochum, Germany, [3] Department of Computer Science, Ruhr-West University of Applied Science, Mülheim an der Ruhr, Germany; [4]Department of Electrical Engineering and Information Technology, Ruhr-University Bochum, Germany; [5]Institute for Experimental Psychophysiology GmbH, Germany

# Corresponding author details:

Omair Ali
omair.ali@ruhr-uni-bochum.de

And

Prof. Dr. Christian Klaes
christian.klaes@gmail.com

# Abstract

Generating continuous electroencephalography (EEG) signals through advanced artificial neural networks presents a novel opportunity to enhance brain-computer interface (BCI) technology. This capability has the potential to significantly enhance applications ranging from simulating dynamic brain activity and data augmentation to improving real-time epilepsy detection and BCI inference. By harnessing generative transformer neural networks, specifically designed for EEG signal generation, we can revolutionize the interpretation and interaction with neural data.

Generative AI has demonstrated significant success across various domains, from natural language processing (NLP) and computer vision to content creation in visual arts and music. It distinguishes itself by using large-scale datasets to construct context windows during pre-training, a technique that has proven particularly effective in NLP, where models are fine-tuned for specific downstream tasks after extensive foundational training.

However, the application of generative AI in the field of BCIs, particularly through the development of continuous, context-rich neural signal generators, has been limited. To address this, we introduce the Generative EEG Transformer (GET), a model leveraging transformer architecture tailored for EEG data. The GET model is pre-trained on diverse EEG datasets, including motor imagery and alpha wave datasets, enabling it to produce high-fidelity neural signals that maintain contextual integrity. Our empirical findings indicate that GET not only faithfully reproduces the frequency spectrum of the training data and input prompts but also robustly generates continuous neural signals.

This research highlights the potential of applying cutting-edge generative AI methodologies into BCI frameworks. By adopting the successful training strategies of the NLP domain for BCIs, the GET sets a new standard for the development and application of neural signal generation technologies.

# 1 Introduction

Generative artificial intelligence (generative AI) is a sub field of AI technology which can produce user specified artificial content including text, images, audio, and synthetic data with the help of generative models [1], [2]. Generative models such as variational autoencoders (VAE) [3] and generative adversarial networks (GANs) [4] have long been at the forefront, allowing for the development of synthetic data samples that exhibit extraordinary diversity and realism [5]. With the advent of diffusion models [6], [7], and the transformer architectures [8], [9], the field of generative AI has revolutionized and made unprecedented progress.

Due to their effectiveness in generating novel and realistic samples, generative AI has found its way in many domains and industries from natural language processing (NLP) [10], [11] to computer vision [12], [13], health care [14], [15], education, research [16], [17] and to art [18]. Initially introduced as a generative model in the domain of NLP, transformers showed that with enough data, it can produce results surpassing other techniques like convolutional neural networks (CNNs). In the NLP domain, pre-trained networks such as Llama 3 and GPT-4 (generative pre-trained transformer) [11], [19] are among the most successful transformer based generative models. The success of generative models in the NLP domain is due to the pre-training of large models (for example GPT-4 and Llama 3) on huge dataset and finetuning them to downstream tasks. Pre-training a model on a big dataset enables it to be utilized as a general-purpose feature extractor, which, after finetuning on downstream tasks with a small dataset, outperforms other models that are typically trained from start for that specific purpose.

Following its success in many other areas, researchers also adopted generative AI in the domain of brain computer interface (BCI) technology. A Brain-Computer Interface (BCI) is a system that allows direct communication between the brain and external devices. By interpreting brain signals, BCIs enable control of computers or machines without physical movement. This technology is crucial for assistive devices, medical rehabilitation, and enhancing human-computer interaction. In the realm of BCI, generative AI is at the frontier of generating synthetic data to augment the training set [20], [21]. Authors in [20] proposed generative model (diffusion model) based data augmentation framework namely Diff-EEG which enables the augmentation of electroencephalography (EEG) signals for detecting Alzheimer's disease. Similarly in [21], authors proposed a framework based on deep convolutional GANs (DCGAN) for generating artificial EEG signals to supplement the training data to improve the classification performance of the classifiers. Generative models such as GANs and VAE have been employed in many studies to enhance the training set to improve the performance of deep learning models for various BCI tasks [22], [23].

Moreover in BCI, generative AI is also gaining significant attention in pattern recognition and predictive modeling tasks [24] and various brain image analysis tasks such as translating neural signals to text (language) or images [25], [26], and segmenting magnetic resonance images (MRI) for identifying traces of lesions after stroke [27]. In [26], functional MRI (fMRI) signals corresponding to neuronal activity in visual and semantic regions of the brain are mapped to image and text components using a generative. Similarly, in [25], a generative model (GPT) is trained to reconstruct semantic language from the fMRI signals. Whereas, in [24], authors proposed a hybrid architecture based on convolution neural network (CNN) and generative architecture (transformer) namely ConTraNet to classify motor imagery signals in EEG data. In [27], the authors developed a brain image segmentation model namely Consistent Perception GAN (CPGAN), which demonstrated higher segmentation performance over other approaches with less labeled data.

Despite the broad success of generative AI in various BCI domains, research on developing generative models for producing continuous, context-based neural signals remains scant. This process is akin to generating ongoing text in NLP tasks, where continuous language output is derived from an input prompt. The ability to create context-sensitive, continuous neural signals is crucial for advancing BCI research across several areas: simulating brain activity, augmenting data with generated neural signals, utilizing the model for real-time prediction of epilepsy, restoring corrupted data from noisy or faulty recordings, and implementing it as a decoder in neural control tasks. We propose that, within the BCI field, pre-training a large-scale model on diverse datasets using the NLP-inspired context window approach will allow the model to understand and reproduce the context-dependent neural signals. This methodology could pave the way for numerous downstream applications. In this work, we present a pipeline based on a generative model (transformer architecture) named Generative EEG Transformer (GET) for pre-training on EEG data and then generating continuous neural signals based on the input prompt or context window.

## 2   Materials and Methods

### 2.1   Data Description

Two different types of neural signals (Motor Imagery (MI) and alpha waves) and two publicly available benchmark datasets are employed to quantify and validate the high generalization capability of the proposed pipeline. These datasets are used for training the generative model GET and employed to provide unseen prompts for neural signal generation. For MI a benchmark dataset containing the data of 9 subjects performing two MI tasks is used [28]. For alpha waves a benchmark dataset from 20 subjects is used [29].

#### 2.1.1   BCI Competition IV dataset 2b (MI-EEG)
The dataset comprises of MI-EEG signals from 9 patients with normal or corrected-to-normal eyesight. Each subject completed five sessions. The MI-EEG data is captured with three bipolar electrodes (C3, Cz, and C4), sampled at 250 Hz, and bandpass filtered from 0.5 Hz and 100 Hz. A notch filter set to 50 Hz is applied. For each trial, the prompts (context input windows) of MI-EEG signal from second 3 to second 5.5 (2.5s in total) containing motor imagining of the hand movement in response to the cue are extracted for training and neural signal generation. The time window corresponding to movement imagination employed here is the same as that used in Ref. [30]. Ref. [28] provides a full description of the dataset.

#### 2.1.2   EEG Alpha waves dataset (alpha-EEG)
This dataset comprises EEG recordings of participants performing a simple resting-state experiment with their eyes open or closed. Data were collected during a pilot experiment held in the GIPSA-lab in Grenoble, France, in 2017. A total of 20 participants participated in the trial.

EEG signals were recorded with EC20 cap with 16 wet electrodes arranged according to the 10-20 international system. The electrode placements were FP1, FP2, FC5, FC6, FZ, T7, CZ, T8, P7, P3, PZ, P4, P8, O1, Oz, and O2. The reference point was located on the right earlobe and the ground at the AFZ scalp region. The data were collected without using a digital filter and at a sampling rate of 512 samples/second.

Every participant participated in a single session consisting of ten blocks of EEG data recording. Each block was ten seconds long. Each subject was asked to record five blocks with eyes closed (condition 1) and the remaining blocks with eyes open (condition 2). The two circumstances were switched around. The participant was instructed to open or close the eyes in accordance with the experimental condition before the start of each block. The detailed description of the dataset is available in Ref. [29].

## 2.2 Methods

Here we present the proposed pipeline of GET for simulating and generating the context based neural signals. As shown in **Figure 1**, the proposed GET consists of three blocks namely: the encoder block, the transformer block and the decoder block.

The encoder block encodes the input context window. It then feeds it to the transformer block, where the attention mechanism extracts the short- and long-term dependencies and learns the relations among the samples of each encoded input. The decoder block uses the attention scored input from the transformer block to generate the output sequence.

**Encoder block:**

The architecture comprises two fully connected linear layers separated by a nonlinear activation function. The first linear layer maps the input signal window into a latent space, serving dual purposes. Primarily, it captures the latent representation of the input signal, essential for extracting vital information necessary for accurate signal representation. Secondly, it facilitates a balance between the breadth of the input signal window and computational efficiency. Expanding the input window enhances the context provided to the model, but also increases computational demands, potentially leading to memory overflow. Conversely, reducing the window size decreases memory usage but at the expense of model performance due to reduced context.

The encoder block addresses this by projecting the input window into a lower-dimensional latent space, thereby managing larger contexts without risking memory overflow by reducing the dimensionality of the input signal window.

The second linear layer operates on the encoded latent space, projecting it linearly into the model dimension $\boldsymbol{d_{model}}$. For this model, the input window spans 150-time stamps, while the latent dimension is set to 100 to prevent memory overflow and minimize computational costs.

A grid search was performed to determine the optimal batch size of 64 and a model dimension $\boldsymbol{d_{model}}$ of 128.

**Transformer block:**

The latent space input signal embeddings, generated by the encoder block, serve as inputs to the transformer block. Initially, positional embeddings, matching the model dimension $\boldsymbol{d_{model}}$, are added to these input embeddings to incorporate positional information. This step is crucial for maintaining the sequential order of events within the neural signal.

The combined embeddings are then processed by the transformer encoder, which comprises seven encoder layers. Each layer features a multi-head self-attention mechanism and a feed-forward block, with layer normalization and residual connections following both components. The multi-head self-attention mechanism is designed to contextually encode the input sequence by calculating the sequence representation across different positions. This method effectively captures the interdependencies among various elements within the same input sequence.

In practical terms, the input embedding vector, combined from the input signal and positional embeddings, is logically divided across multiple 'heads' in the self-attention layer. This division allows each segment of the embeddings to concurrently learn different facets or aspects of the input sequence in relation to other sequence elements. Such parallel processing leads to a more nuanced and detailed representation of the input sequence, significantly enhancing the transformer encoder's ability to capture rich representations.

The transformer encoder layer consists of 6 encoder layers and each multi-head self-attention layer uses 8 heads resembling the original transformer architecture [8]. A feed forward dimension of 256 is used in the feed-forward block of each encoder layer. Dropout is employed to avoid overfitting during training.

**Decoder block:** Similar to the encoder, the decoder block consists of two fully connected linear layers, separated by a nonlinear activation function. The first linear layer in the decoder block transforms the encoded output from the transformer block, mapping it from the latent space dimension back to the original input space dimension. This adjustment ensures that the output sequence length matches the input sequence length. The second linear layer then processes this transformed input, converting it from the model dimension $\boldsymbol{d_{model}}$ to the output feature dimension.

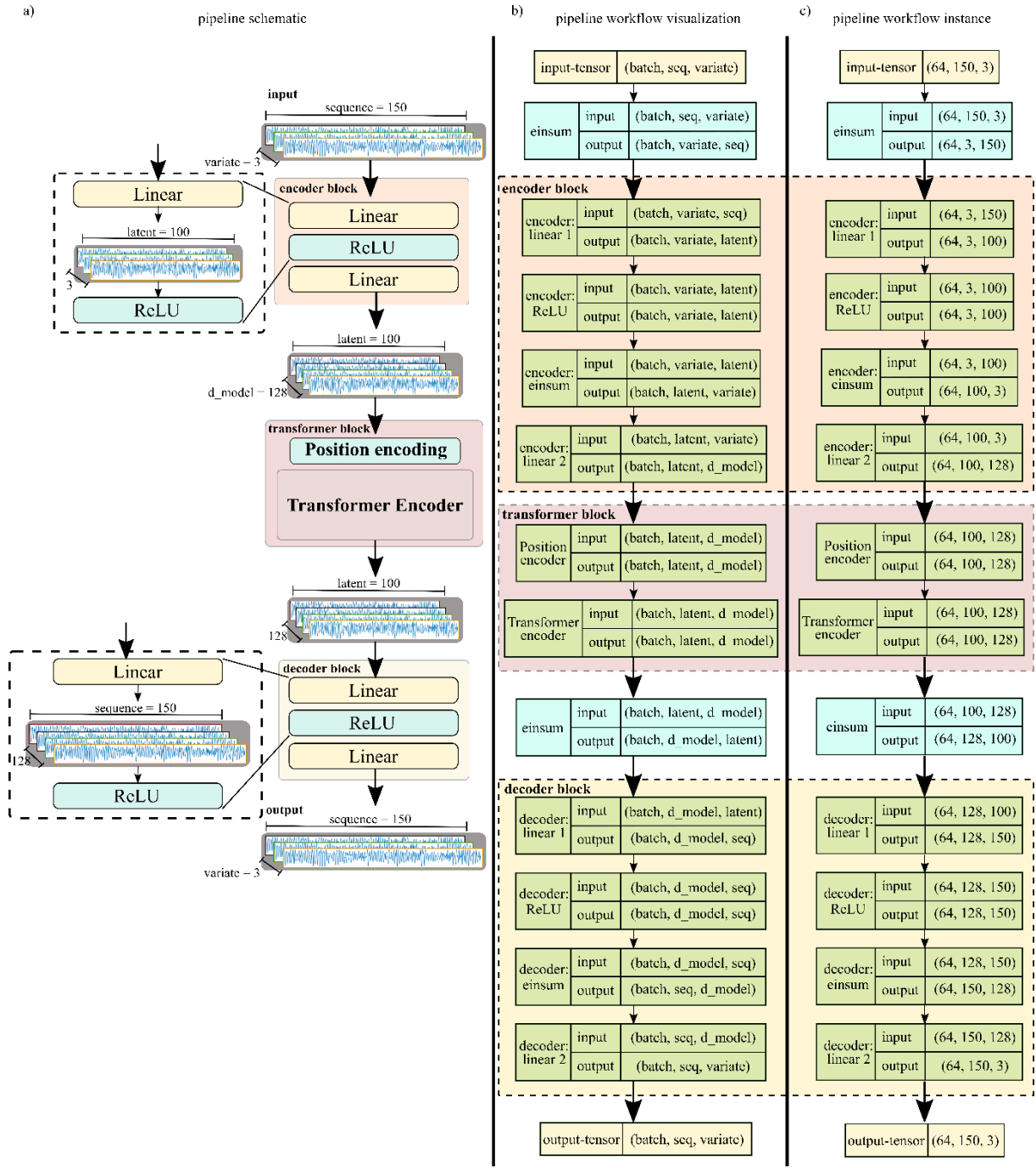

*Figure 1: GET. Pipeline for context based neural signal generation. It has three blocks: an encoder block, a transformer block and a decoder block. Input is first fed to encoder block which projects it to latent space and then its embeddings are learned by projecting it to model dimension. The output of the encoder block is the input to the transformer block which learns the relationship between different elements of its input. The output of the transformer block is used as input of the decoder block which maps the input from latent space to the original dimension.*

**Workflow of information through pipeline:** The information workflow through GET as shown in **Figure 1 (b** and **c)** is as follows:

- **Input:** The input tensor (with dimensions batch, variate and sequence) is fed to the encoder block. Where the variate represents the number of variables (number of electrodes) present in a signal. Univariate refers to signal from single electrode whereas multivariate refers to signals from multiple electrodes. Sequence is the length of input signal which is equal to the input signal window and batch represents the number of samples in single input.
- **Encoder block: (Linear 1):** First fully connected linear layer operates on sequence dimension of the received input tensor and projects it to a latent space with new tensor (of dimensions batch, variate and latent).
- **Encoder block: (ReLU):** Non-linear activation is applied to the output of first encoder linear layer and outputs the same dimensional tensor (batch, variate, latent).
- **Encoder block: (Linear 2):** Second fully connected linear layer learns the embeddings of each sample by projecting it from variate dimension to model dimension $d_{model}$. The resultant tensor (batch, latent, $d_{model}$) is fed to the transformer block.
- **Transformer block (Position encoding):** Fixed position encodings of dimension $d_{model}$ resembling the original transformer architecture [8] are added to the received tensor from the encoder block. The resulting tensor (batch, latent, $d_{model}$) is used as an input to the transformer encoder layer.
- **Transformer block (Transformer encoder):** After coupling the position information with the learned embeddings, the resultant tensor of shape (batch, latent, $d_{model}$) is fed to the transformer encoder. Therefore, the transformer block outputs the tensor of shape (batch, latent, $d_{model}$). The details of the functionality of transformer in general and self-attention in particular can be found in [8].
- **Decoder block: (Linear 1):** The output of the transformer block is fed to the decoder block. The first linear fully connected linear layer projects the latent dimension of its input to match the sequence dimension of original input-tensor, thus resulting in the tensor (batch, $d_{model}$, sequence).
- **Decoder block: (ReLU):** Like the encoder ReLU activation layer, this adds a non-linearity to the output and results in a tensor (batch, $d_{model}$, sequence).
- **Decoder block: (Linear 2):** The second fully connected linear layer operates on the model dimension $d_{model}$ and learns to map from model dimension to output variate dimension. As a results, it produces a tensor (of dimensions batch, sequence and variate).

# 3 Results

Here, we present a detailed performance evaluation of our proposed pipeline for generating context-based neural signals that simulate brain activity. Our analysis comprises two main methods. First, we analyzed the Fourier spectrum of both the generated signal and the entire training dataset to determine if the trained model accurately learned and replicated the frequency spectrum. Second, we employed the Short Time Fourier Transform (STFT) to assess whether the generated signal captured the context of the prompt through its temporal variations in frequency.

Given the challenges of learning from neural signals—namely their low signal-to-noise ratio and non-stationary nature—we initiated our evaluations with univariate signal generation, subsequently advancing to multivariate neural signals.

### 3.1   Univariate neural signal generation

Here, we present the ability of our proposed pipeline in learning the context and generating the corresponding univariate neural signal. In this analysis, we first employed the filtered signals to train the pipeline. We then fed the unseen prompt to the trained model which generates the neural signal based on the prompt. In this experiment we jointly trained the model on two different EEG datasets (MI-EEG and alpha-EEG dataset). Each dataset contributed equally during the training of the model. In the second step, we used the raw neural signals to train the model.

**Univariate filtered signal generation: Figure 2** shows three instances (a, b, and c) of univariate signal generation based on three randomly selected unseen prompts. In **Figure 2**, each column represents an instance or an example. The first row represents the prompt (in blue) and the generated signal (in red). Second and third rows show the frequency spectrum of the generated signal and the entire training data respectively. In this experiment, we used input windows of size 100 samples to train the model, whereas the generated signals have length equal to 600 samples. Rows two and three clearly indicate that, the generated signals simulate the frequency spectrum of the training data thus indicating the learning of frequency distribution.

Similarly, **Figure 3** represents another three instances of univariate signal generation based on randomly selected unseen prompts. Here, these three instances are taken from alpha waves dataset. It is evident from **Figure 3**, that the produced signals follow the frequency distribution of the alpha-EEG dataset. However, in both cases, the frequency amplitude of the generated signals is slightly higher than the average frequency amplitude of the respective signals. One reason could be the that since we used the latent representations of the input signals to train the model to avoid the memory overflow, it lost some information during this transformation. Secondly, since we employed only the mean squared error as the loss function, it becomes non-trivial to optimize for frequency amplitude without customizing the loss function to include the frequency component.

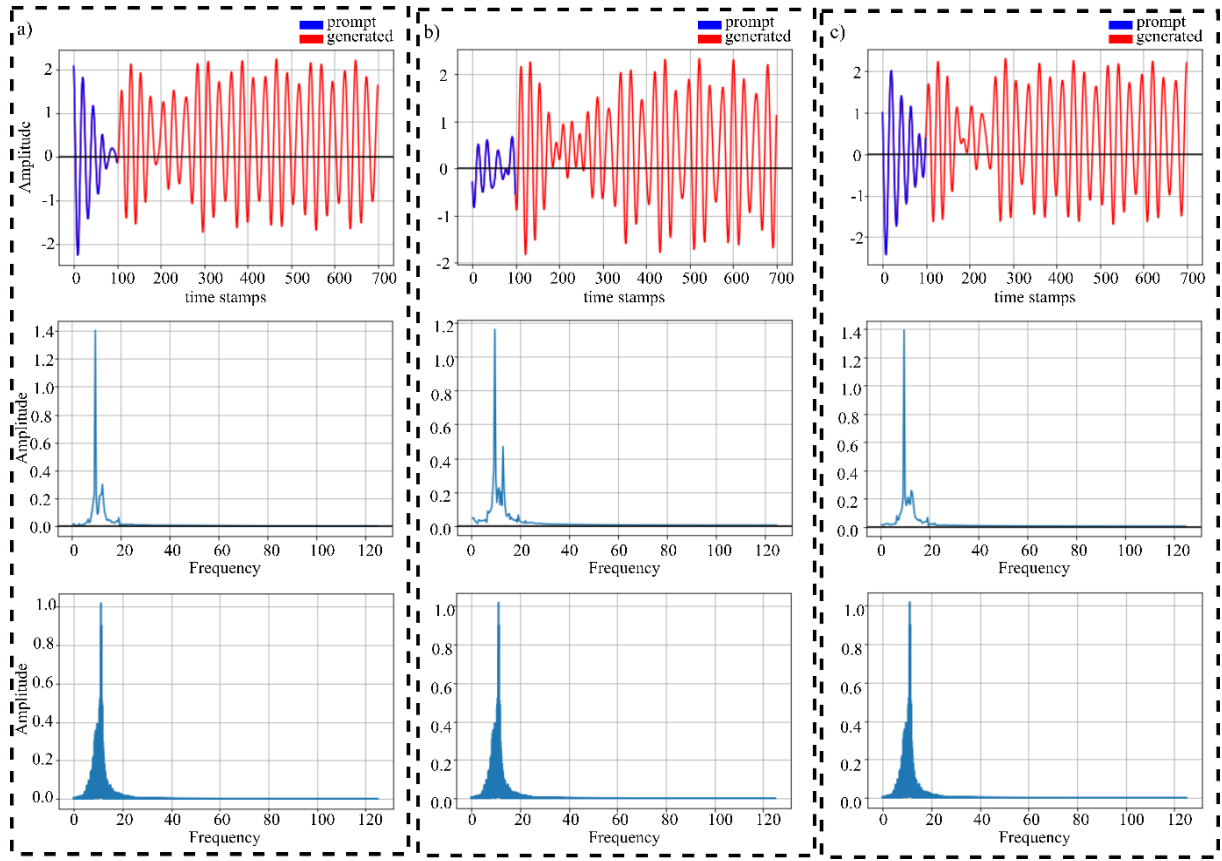

*Figure 2:* *Univariate signal generation of motor imagery EEG signals with prompt of 100 samples. a), b) and c) show three examples of neural signal generation based on randomly selected unseen three input prompts from MI-EEG dataset. For each example, first row shows the input prompt in blue and generated signal in red color. Second row shows the frequency spectra of generated signal whereas the third row shows the frequency spectra of the training data.*

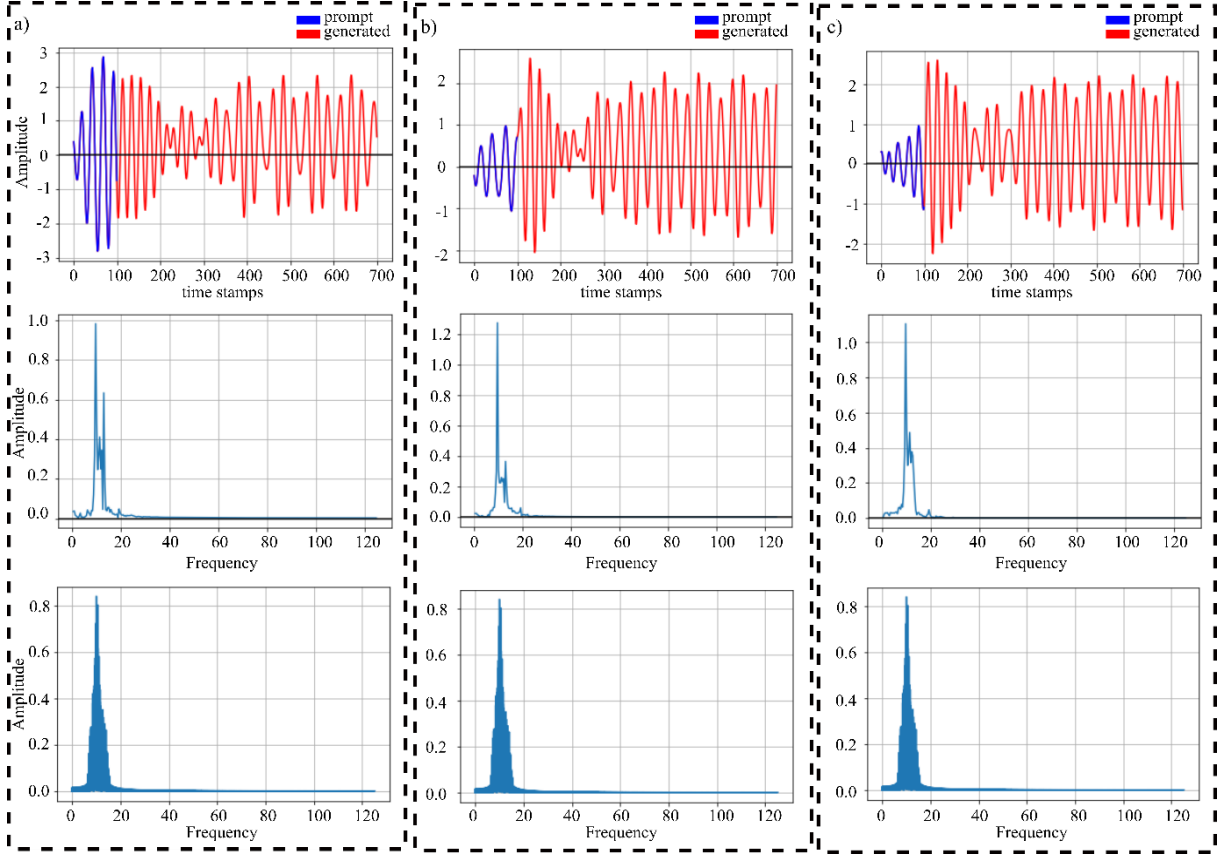

*Figure 3: Univariate signal generation of alpha waves neural signals with prompt of 100 samples. a), b) and c) show three examples of neural signal generation based on randomly selected unseen three input prompts from alpha waves dataset. For each example, first row shows the input prompt in blue and generated signal in red color. Second row shows the frequency spectra of generated signal whereas the third row shows the frequency spectra of the training data.*

### 3.1.1 Statistical analysis

In order to find the statistical significance of the generated signals by GET, we computed the power spectral densities (PSD) of the generated signals and compared them with the corresponding ground truth signals. Moreover, we also computed the mean squared error (mse) of the power spectral densities of generated and ground truth signals. PSD is the measure of power content of the signal over its frequency spectrum. It helps to understand how closely the power distribution of the generated signal corelates with that of the ground truth signal. **Figure 4** shows the PSD comparison between the generated and ground truth signals based on three unseen prompts for MI-EEG (a1, a2 and a3) as well as alpha-EEG signals (b1, b2 and b3). It is shown in **Figure 4**, that the signals generated by GET closely follow the power spectral distribution of the ground truth signals indicating the ability of the GET to learn the context from the training data. The mse between the PSD of generated and ground truth signals as reported in **Table 1** quantifies this correlation.

*Table 1: Mean squared error (mse) between PSD of generated and ground truth signals.*

| Mean squared error (mse) | a1) 0.0273 | a2) 0.0354 | a3) 0.0132 |
|---|---|---|---|
| | b1) 0.0034 | b2) 0.0029 | b3) 0.0038 |

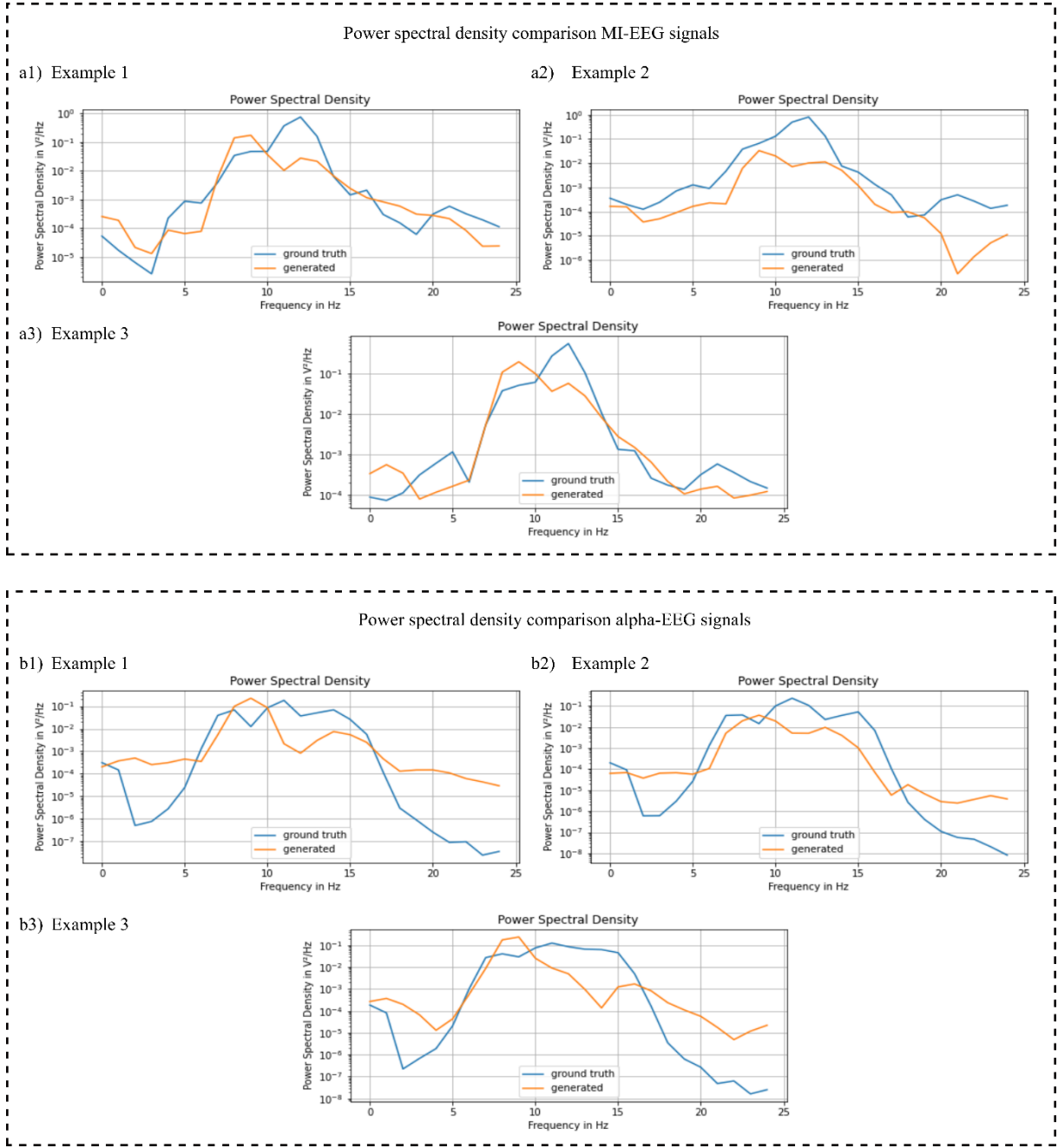

***Figure 4:*** *Power spectral density comparison between generated and ground truth signals of MI-EEG signals and alpha-EEG signals. a1,2 and 3) represent the comparison between power spectral densities of generated and ground truth signals based on three randomly selected prompts of MI-EEG signals. b1, 2 and 3 represent the comparison between power spectral densities of generated and ground truth signals based on three randomly selected prompts of alpha-EEG signals.*

In another experiment, we jointly trained the model on both datasets using a longer input window with size of 200 samples. In this experiment, both datasets contributed equally during the training process. Here since the context window is doubled compared to previous experiment, we allowed the model to generate longer sequences of size 1000 samples as shown in **Figure 10**.

**Figure 10** presents three instances (a, b, and c) of signal generation based on three randomly selected prompts from normal EEG dataset whereas the instances (d, e, and f) show the signal generation based on prompts taken from alpha waves dataset. It is shown in **Figure 10** (a, b, and c) that model captures the frequency distribution of the normal EEG dataset. However, in this case it also picks up some small magnitude of noise belonging to higher frequency ranges such as between 45-50Hz and 80Hz. Similarly, it is shown in **Figure 10** (d, e, and f) that the model learns the frequency distribution of the alpha waves dataset however, it also produces very small magnitude of noise components in higher frequencies. The noise produced by alpha wave signal generation is smaller in magnitude compared to noise produced by normal EEG signal generation. One probable reason for the generation of noise component is that for longer signal generation, the error accumulates over time which results in these noise components. To mitigate this effect, we retrained the model with higher regularization during training. The result is presented in **Figure 11**. **Figure 11** (a) shows the instance of long sequence generation based on prompt of normal EEG signal, whereas **Figure 11** (b) indicates the signal generation based on alpha wave prompt. In this experiment, we allowed the model to generate even longer sequence of length 1500 samples. Both examples of **Figure 11** show that model learned the frequency distribution of the datasets and the regularization aided to curtail the effect of noise components while producing even longer sequences.

**Univariate unfiltered signal generation:** In order to evaluate if the generated signals captured the context of the prompt, we performed time frequency analysis to analyze the temporal variations in the frequency spectrum. Henceforth, in this experiment, we computed the spectra of the prompt as well as the generated signal using STFT algorithm. **Figure 5** presents the five instances of neural signal generation based on five randomly selected prompts. Where, each row presents an example of neural signal generation based on the prompt. Second column shows the spectra of the given prompts whereas the third column depicts the spectra of the generated signals. In this and the following experiments employing unfiltered signals, we trained the models using the prompt size of 150 samples which was obtained empirically. Increasing the prompt size in this case adds more context which is prone to more noise which consequently makes it non-trivial to project to latent representation which is then used by the transformer encoder for training. Henceforth, we found empirically the optimal size of input window with 150 samples in case of unfiltered signals.

The prompt in the example displayed in **Figure 5** (a) includes activities in the frequency range of 5-8 Hz, 10-12 Hz, and some activity near 20 Hz. The signal that is generated adheres to the prompt's context and produces comparable activities in those frequencies, along with some extra activity in the sub-5 Hz region. Comparably, in the example shown in **Figure 5** (b), when the prompt exhibits dominant activity in the 10 Hz range, the generated signal also exhibits activity in the 10 Hz region, but it also exhibits activity in the 5-6 Hz range. Given the nearly 2.5-second duration of the generated signal, it is anticipated to elicit activity in other frequency ranges in order to replicate the realistic brain signals.

Similarly, in the example shown in **Figure 5** (d), where the prompt elicited major activities in range 10-12 Hz, the produced neural signal also depicts the activity in this range. However, it also generates strong activities in ranges 5-8 Hz which concords with the activity ranges of the entire training data.

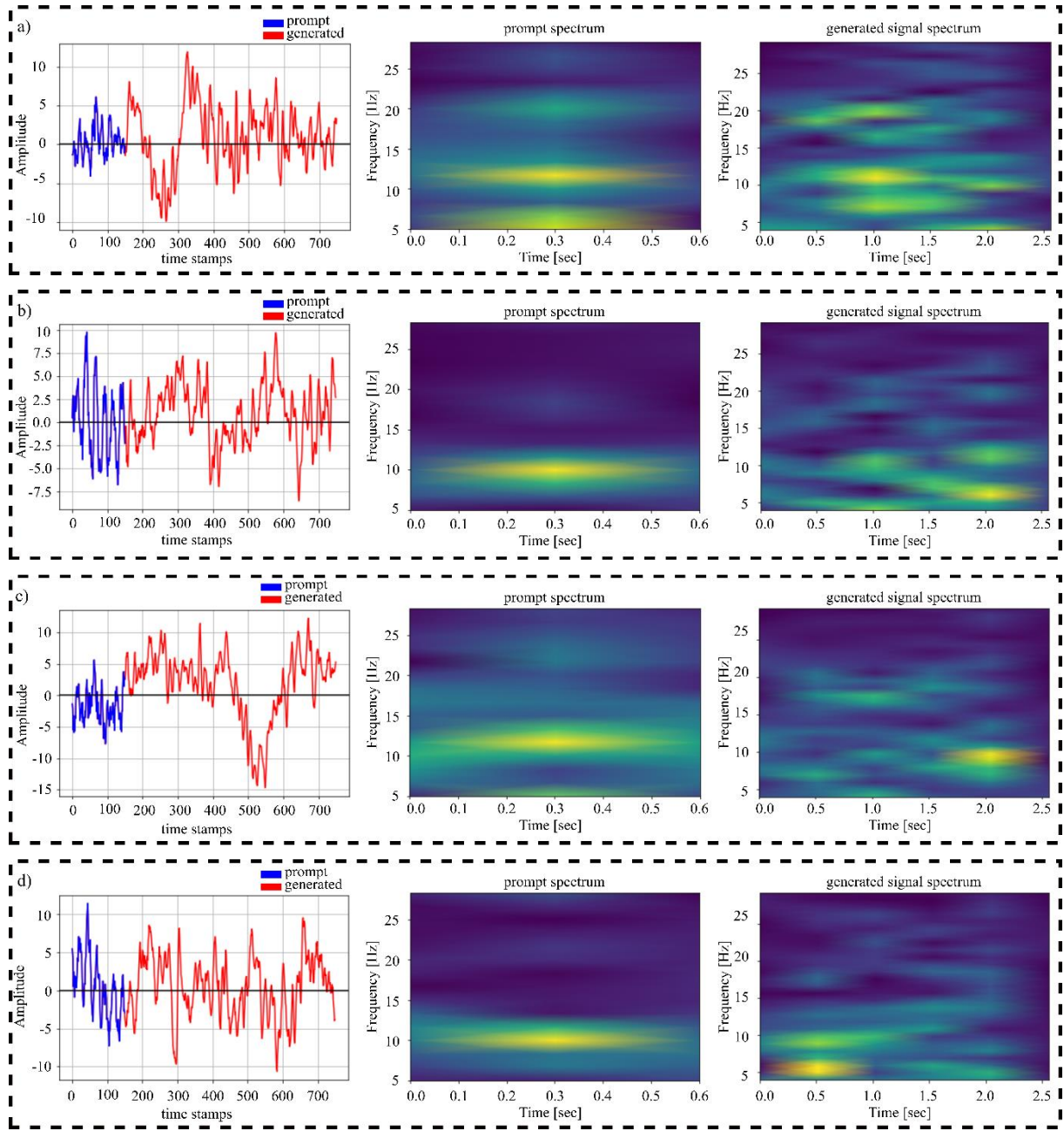

***Figure 5:*** *Univariate signal generation of neural signals with unfiltered prompt of 150 samples. a) and b) show two examples of neural signal generation based on randomly selected unseen input prompts of MI-EEG dataset. Whereas c) and d) show two examples of neural signal generation based on randomly selected unseen input prompts of alpha waves dataset. For each example, first column shows the input prompt in blue whereas the generated signal in red color. The second column shows the frequency spectra of the input prompt generated by STFT whereas the third column represents the frequence spectra of generated signal.*

## 3.2 Multivariate neural signal generation

Here, we demonstrate the scaling ability of our proposed methodology to generate multivariate neural signals simultaneously. For this purpose, we extended the number of electrodes from 1 to 3. The model is trained in similar fashion as in univariate case except for output layer since in multivariate case, it regresses more than one value simultaneously.

**Figure 6** shows an example of multivariate neural signal generation. Unlike univariate signal generation, each row here corresponds to neural signal generation of an electrode. Since in this case, we have 3 electrodes, thus three rows. Column 2 shows the spectra of the given prompt of an electrode, whereas column 3 represents the spectra of the generated signals of the respective electrodes.

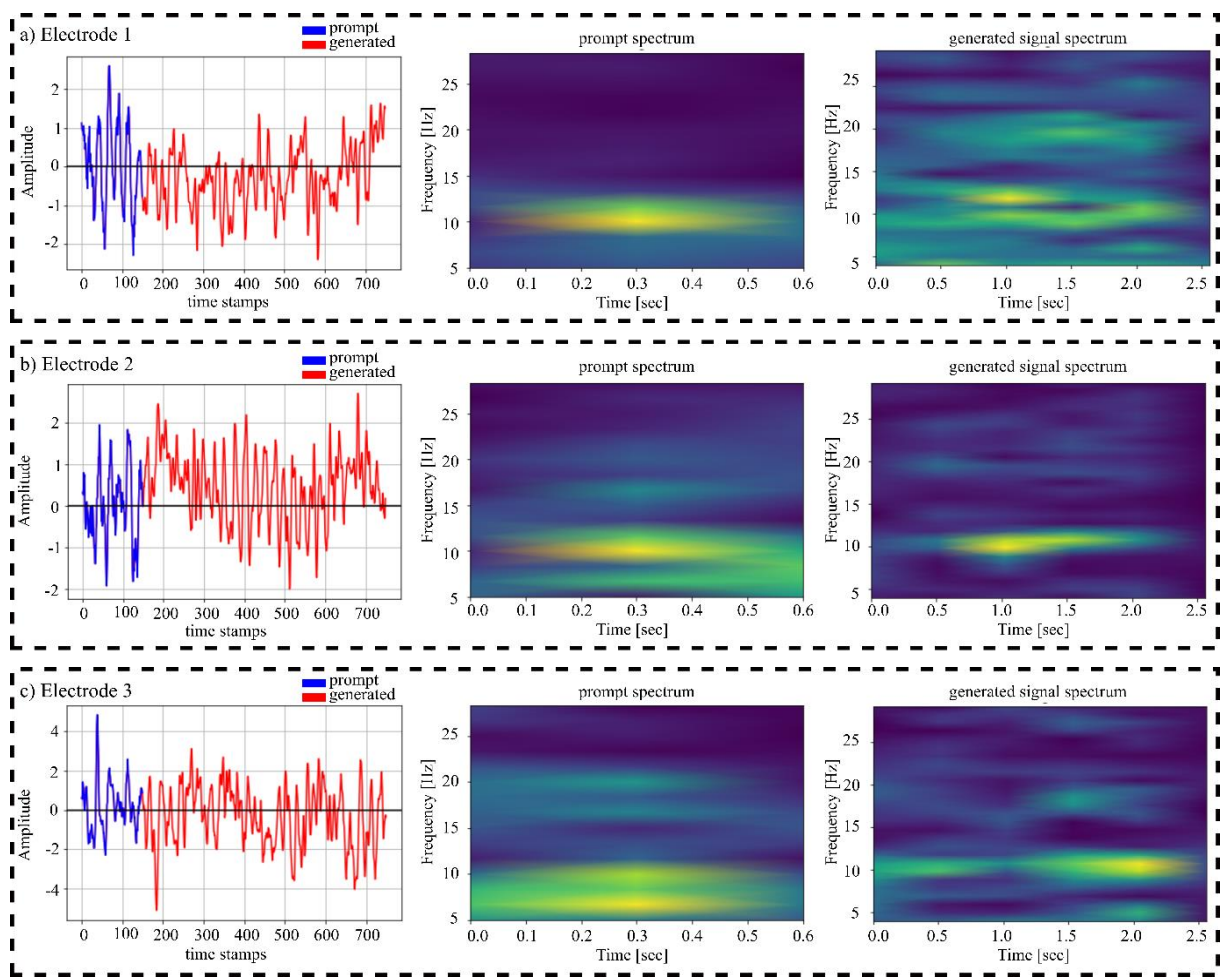

*Figure 6: Example 1. Multivariate signal generation of neural signals with unfiltered prompt of 150 samples. a), b) and c) show an example of the multi-electrode neural signal generation based on the unseen prompt from alpha waves dataset. Here each row corresponds to the neural signal generation of the respective electrode. For each electrode, the first column represents the input prompt in blue whereas the generated signal in red color. Columns two and three show the frequency spectra of input prompts and the generated signals respectively.*

It is shown in **Figure 6** (a) that prompt of electrode 1 triggered activity in range 10-12 Hz. The corresponding generated signal captures the context of the prompt and produces the similar activity in that range. However, it additionally produces activities in range 12-13 Hz and 20-

22 Hz. However, in case of electrode 2 as shown in **Figure 6** (b), the generated signal captured the dominant activity range of the prompt which is around 10 Hz and elicited a strong activity in that range. Contrarily, in case of electrode 3 as presented in **Figure 6** (c), where the prompt triggers the activity between 5-10 Hz, the generated signal produced dominant activity near 10 Hz and mild activities near 5 Hz.

In another example shown in **Figure 12,** generated signals capture the context of the respective prompts and produce similar activities. More specifically, prompt of electrode 1 (**Figure 12 (a)**) elicited dominant activity in near 10 Hz which is successfully represented by the model as the generated signal also contains the dominant activity near 10 Hz. Similarly, the prompt of electrode 2 (**Figure 12 (b)**) shows main activity near 10 Hz which is replicated by the neural signal generated in accordance. Furthermore, it also captures the less prominent activity ranges of the prompt including those between 5-8 Hz and produces similar activity in that range. However, the prompt of electrode 3 (**Figure 12 (c)**) elicited activities between 8-10 Hz as well as near 20 Hz. Similar activity is seen in the spectra of the respectively generated neural signal indicating that model captured the context of the prompt.

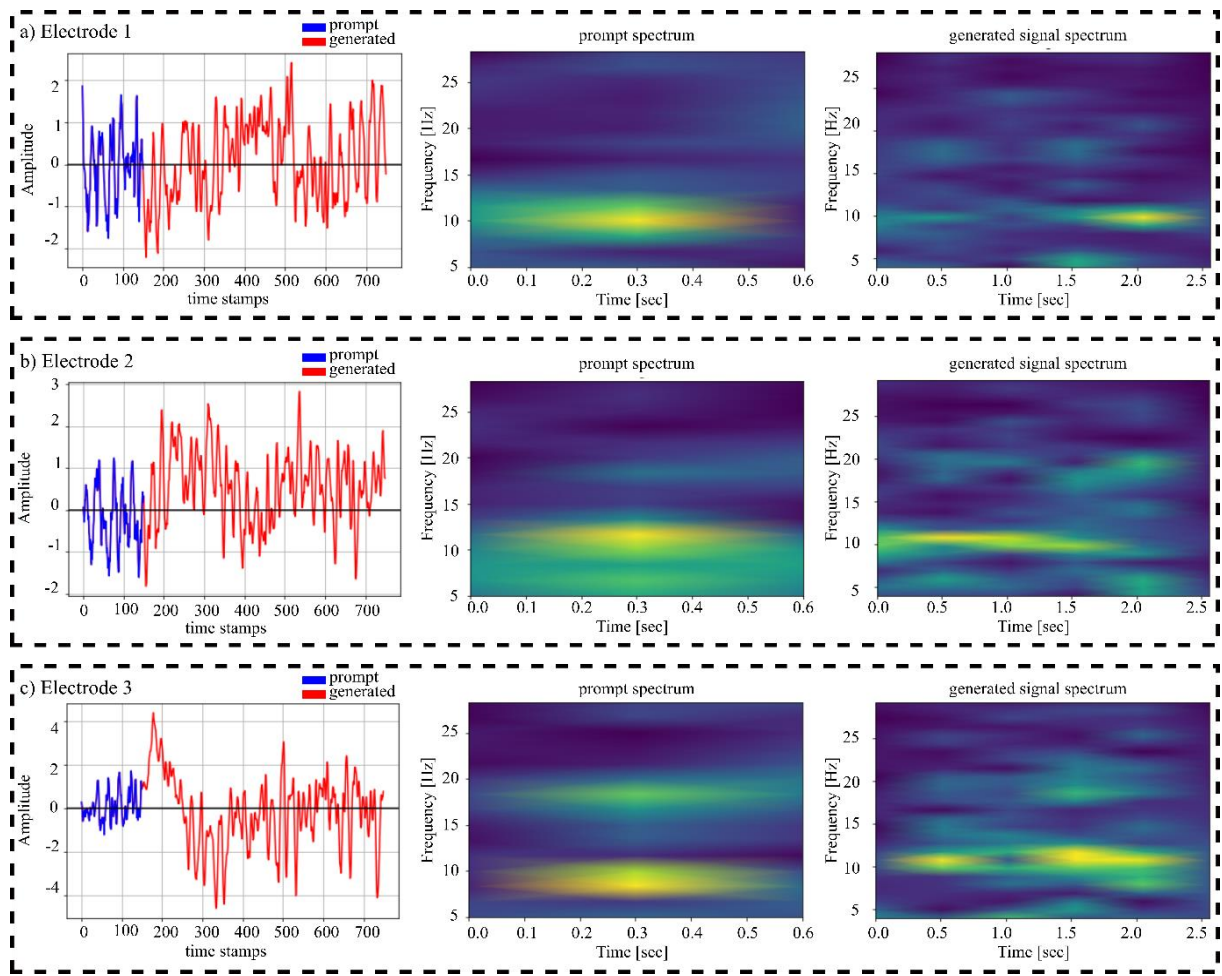

*Figure 7: Example 2. Multivariate signal generation of neural signals with unfiltered prompt of 150 samples. a), b) and c) show an example of the multi-electrode neural signal generation based on the unseen prompt from alpha waves dataset. Here each row corresponds to the neural signal generation of the respective electrode. For each electrode, the first column represents the input prompt in blue whereas the generated signal in red color. Columns two and three show the frequency spectra of input prompts and the generated signals respectively.*

**Figure 7** shows another example of multivariate neural signal generation, where the model learned to extract the prompt's context and generate neural signals in accordance. Prompts of electrodes 1 and 2 (**Figure 7** (a and b)) generate activities in frequency range 8-12 Hz with more dominant activities close to 10-12 Hz. The generated signals mirror this behavior and produce strong activities near 10 Hz. In case of prompt of electrode 3 (**Figure 7** (c)), the prominent activity lies in range 17-19 Hz and less prominent activity is produced near 10 Hz. This behavior is mirrored by the generated signal as it also elicits dominant activity in range 17-19 Hz and slightly less dominant activity near 10 Hz. This clearly indicates the ability of the learned model to understand the context of the given prompt.

Similarly, another instance of neural signal generation based on context is depicted in **Figure 13** Here, the prompts of electrode 1 and electrode 2 (**Figure 13** (a and b)) produced prominent activities in range 8-12 Hz which is mirrored by the respective generated signals. The generated signals based on the prompts of electrodes 1 and 2 as shown in **Figure 13** (a and b) also elicited fewer dominant activities in similar range of given prompts. However, for electrode 3, the prompt contained activities in ranges 10-12 Hz as well as near to 20 Hz. The activity pattern is well extracted by the model, and it generated the signal by replicating those activities in the similar ranges. Moreover, it also elicited activities in sub 5 Hz ranges which concord with the frequency spectrum of the training data.

### 3.2.1 Comparison with ground truth and statistical analysis

In order to evaluate the context learning capability of the GET for multivariate case, it is necessary to find the statistical significance of the generated signals and compare them with the ground truth signals. Henceforth, here we performed the statistical analysis and computed the PSD of the generated signals and compared them with the PSD of the ground truth signals. **Figure 8** shows the comparison between the generated signals and the ground truth signals. Here a1, b1 and c1 show the generated signals and their corresponding spectra based on the given prompts whereas a2, b2 and c2 show the corresponding ground truth signals and their spectra based on the same prompt. It is evident from the **Figure 8**, that signals generated by GET follow the same frequency distribution as that of the ground truth signals. More concretely, in **Figure 9**, PSD comparison between the generated and ground truth signals is presented. Here the PSD is computed by two different methods namely Periodogram and Wlech's method to get the raw as well as the smooth version of the PSD as shown in a1, b1 and c1. The PSD comparison is shown in a2, b2 and c2. It is evident from the figure that the PSD of the generated signals follow the pattern of the ground truth signals indicating the context learning ability of GET model. To further quantify these results, we also computed the mse between the PSD of generated and the ground truth signals which is shown in **Table 2**

*Table 2: Mean squared error (mse) of the PSD of ground truth and generated signal.*

| mse | a2) 0.000656 | b2) 0.000644 | c2) 0.005493 |
|---|---|---|---|

Similarly, another example depicted in **Figure 14** shows the comparison of the generated signals and the corresponding spectra with ground truth signals and their spectra. The statistical analysis as shown in **Figure 15** also represent that the signals generated by GET follow the power distribution of the ground truth signals. This correlation is further quantified by computing the mse between PSD of generated and ground truth signals which is shown in **Table 3**.

The results reported in this, and the aforementioned section shows the ability of our proposed pipeline to learn and extract the context from the prompt and generate signal in accordance.

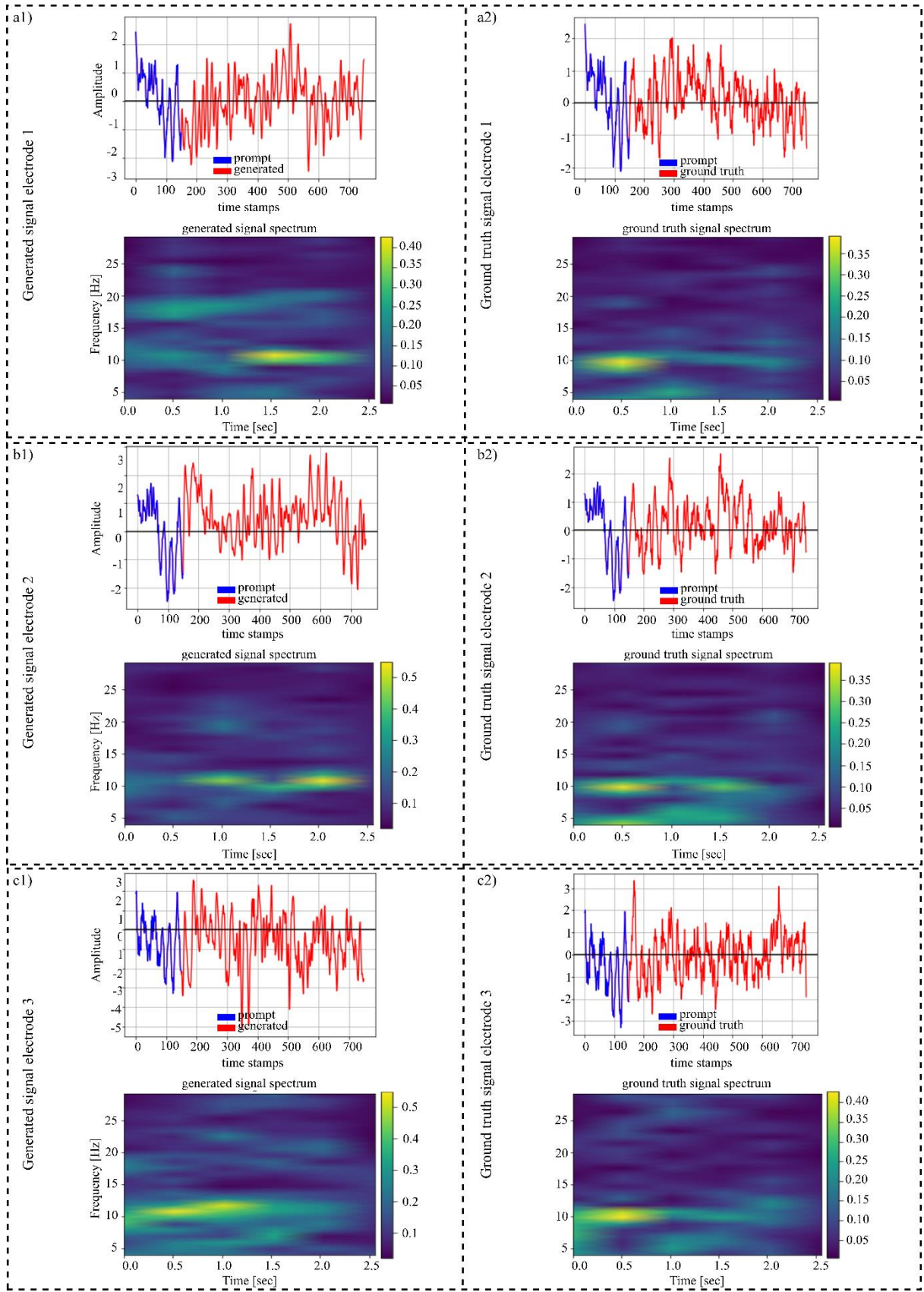

*Figure 8:* Comparison between generated signals and the ground truth signals based on unseen prompts. a1, b1 and c1) represent the signals generated by GET and their corresponding spectra whereas a2, b2 and c2) represent the ground truth signals and their corresponding spectra.

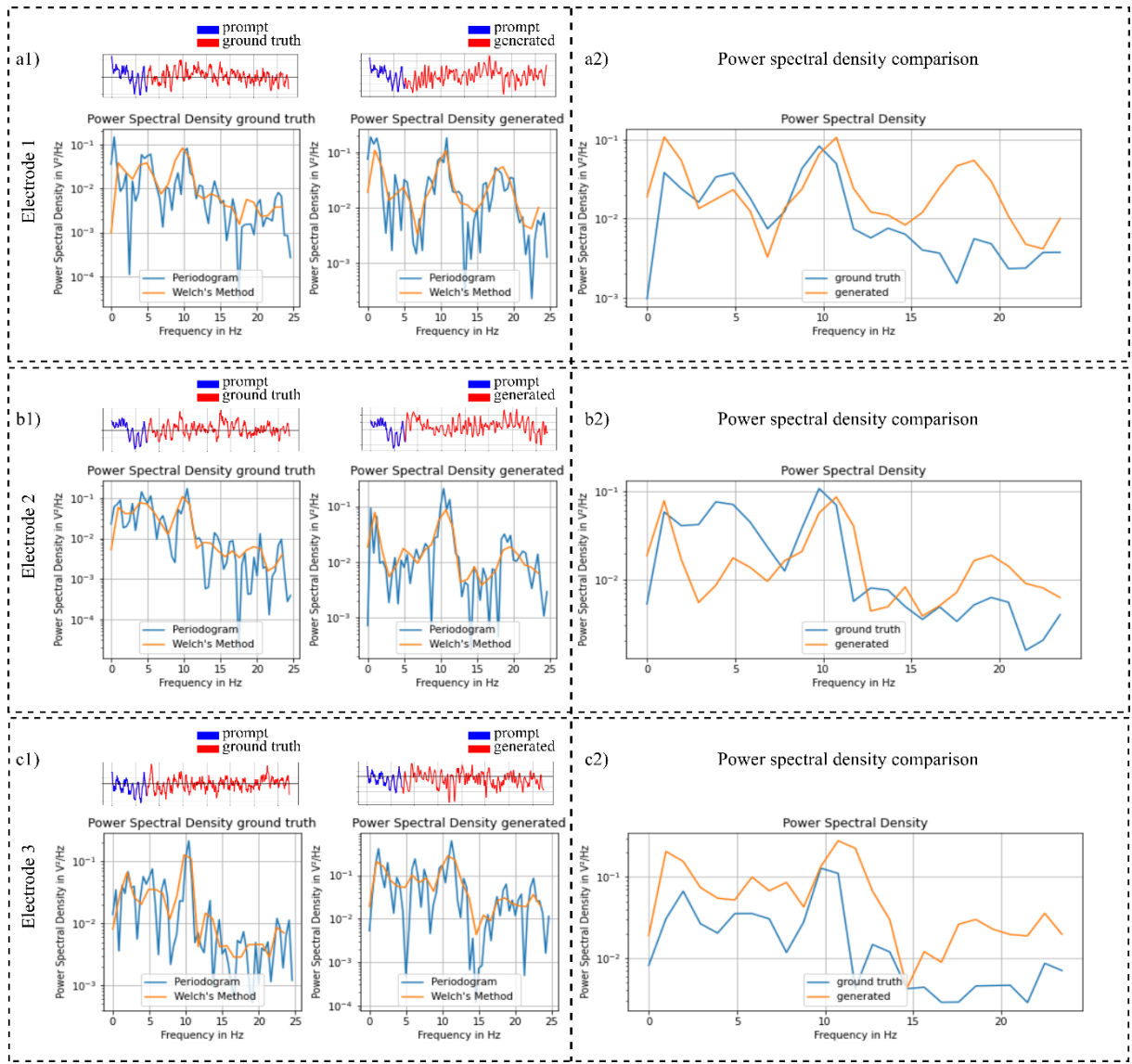

***Figure 9:*** *PSD of the generated signals and ground truth signals based on the given prompt. a1, b1 and c1) show the power spectral densities of the generated and ground truth signals computed using two different methods. a2, b2 and c2) show the comparison between the PSD of generated and ground truth signals.*

# 4  Discussion

In this work, we attempted to generate artificial neural signals based on input prompt using proposed GET model. Proposed model successfully learns the context from the prompt (incomplete input) and generate the neural signal similar to the original sample. **Figure 8** shows the input prompt and the complete ground truth EEG signal in **Figure 8** (a2, b2 and c2), and the in **Figure 8** (a1, b1 and c1), shows the corresponding generated EEG signals. Original time series signal and the generated time series signal along with their spectrogram and the power spectral densities show the high generalization quality of GET.

The proposed pipeline of GET consists of three modules namely encoder block, transformer block and a decoder block. One of the intuitions behind employing the stand-alone encoder and decoder block in addition to transformer block is to address the trade-off between performance and memory constraints.

We also evaluated the quality of the generalization of the proposed model based on the length of input prompt. Larger input prompt windows enrich the model (see **Figure 10** and **Figure 11**) with more available context to learn from at the cost of computational expensiveness. On the contrary, feeding smaller input windows may lack the context to extract the meaningful information from them. It consequently can affect the ability of the model to generate the signals in accordance with the prompt. This trade-off is addressed by learning the latent representations of the larger input prompt windows (150 samples) by projecting them to latent space with fewer dimensions (100 samples). The decoder at the end transforms the latent space representation (100 samples) back to its original form (150 samples) thus resulting into generated signal. While the transformer block in between learns the long as well as short term dependencies present in the input signal.

We evaluated the proposed pipeline for generating univariate filtered as well as unfiltered signal based on the given prompt. For this experiment, we trained the model jointly on MI-EEG and alpha-EEG dataset. The results presented in **Figures (2, 3, 4, 5, 10 and 11)** show that the model learned to extract the context from the training data and generalized well on the unseen prompts.

The signals generated based on unseen prompts indicate that the model learned the frequency spectrum from the training data as the frequency spectrum of the generated signals concords with that of the training data. Similarly, the STFT spectra of the unfiltered generated neural signals as shown in **Figure 5** show that the model mirrors the frequency spectrum of the given prompt as well.

Moreover, we also evaluated the GET for generating multivariate neural signals based on the input prompt. In this case, we trained the GET on joint data distribution of MI-EEG and alpha-EEG dataset to generate context based neural signals for three electrodes. The results depicted in **Figures (6, 7, 8, 9, 12 and 13)** show the capability of GET to maintain the context window to learn the context and produce the respective continuous neural signals. As for alpha-EEG signals, where the expected operating frequency is around 8-10Hz during eyes closed task, the model picked the context from its pre-training and generated signals with similar operating frequency as presented in **Figure 6** and **Figure 12**. Similarly for MI-EEG dataset, GET learned the frequency spectrum from the training data and generated signals in accordance as shown in **Figure 7** and **Figure 13**

We believe that the ability to generate continuous, context-sensitive neural signals is essential for furthering research in BCI in several areas such as brain activity simulation, data augmentation with generated neural signals, epilepsy prediction in real time using the model, data restoration from damaged or noisy recordings, and application of the model as a decoder in neural control tasks. To the best of our knowledge, we have proposed the first generative machine learning model for generating neural signals based on the given neural prompt. This work has the potential to lay the foundation to solve many aforementioned research challenges faced by BCI community.

# 6 Appendix
## 6.1 Univariate neural signal generation

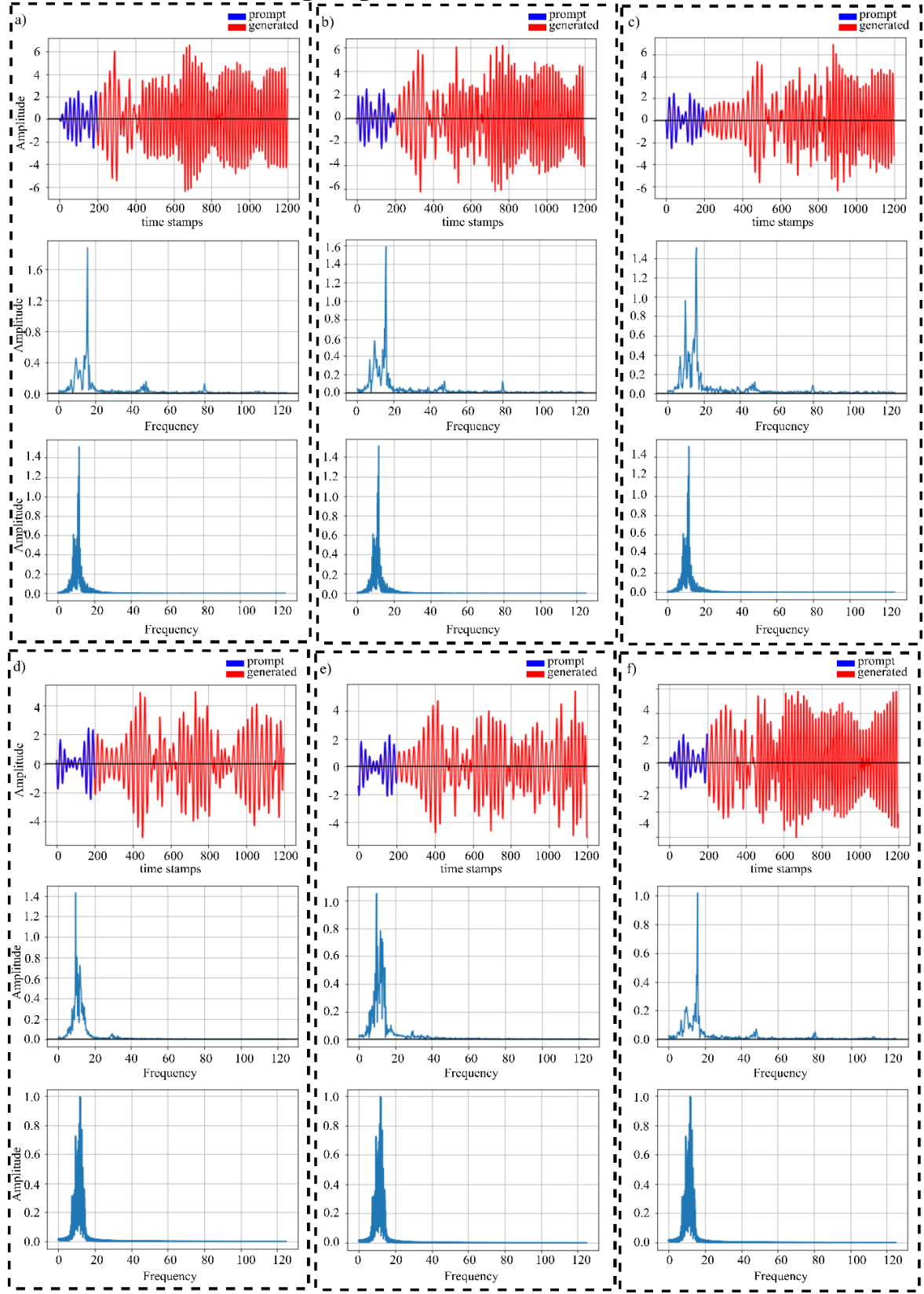

*Figure 10:* Univariate signal generation of MI-EEG and alpha -EEG signals with prompt window of 200 samples. a), b) and c) show three examples of neural signal generation based on randomly selected unseen three input prompts of MI-EEG

*dataset. Whereas d), e) and f) show three examples of neural signal generation based on randomly selected unseen three input prompts of alpha-EEG dataset.*

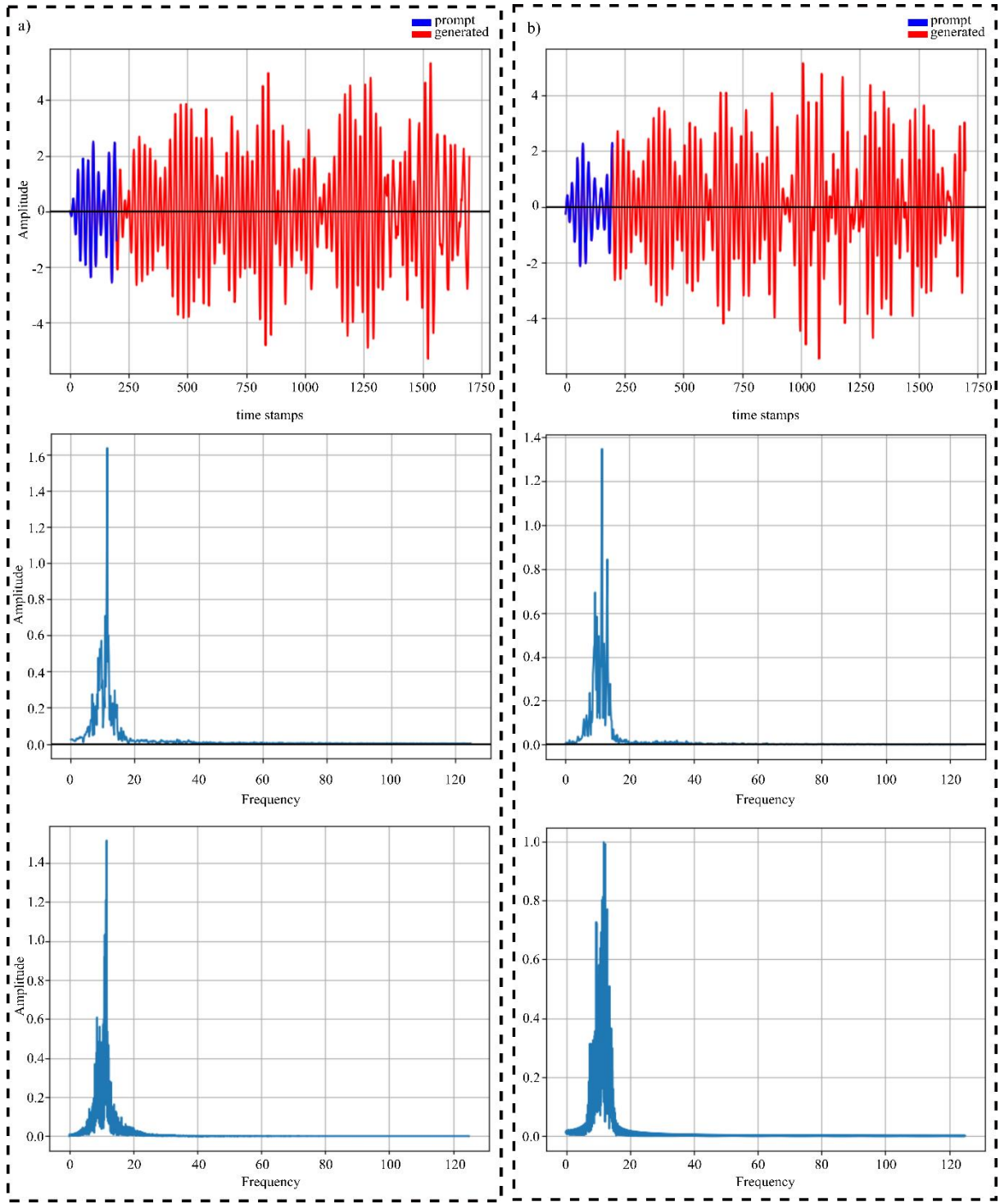

*Figure 11: Univariate long sequence signal generation of MI-EEG and alpha-EEG neural signals with prompt window of 200 samples. a) and b) show examples of neural signal generation based on randomly selected unseen input prompts of MI-EEG dataset and alpha-EEG dataset respectively.*

## 6.2 Multivariate neural signal generation

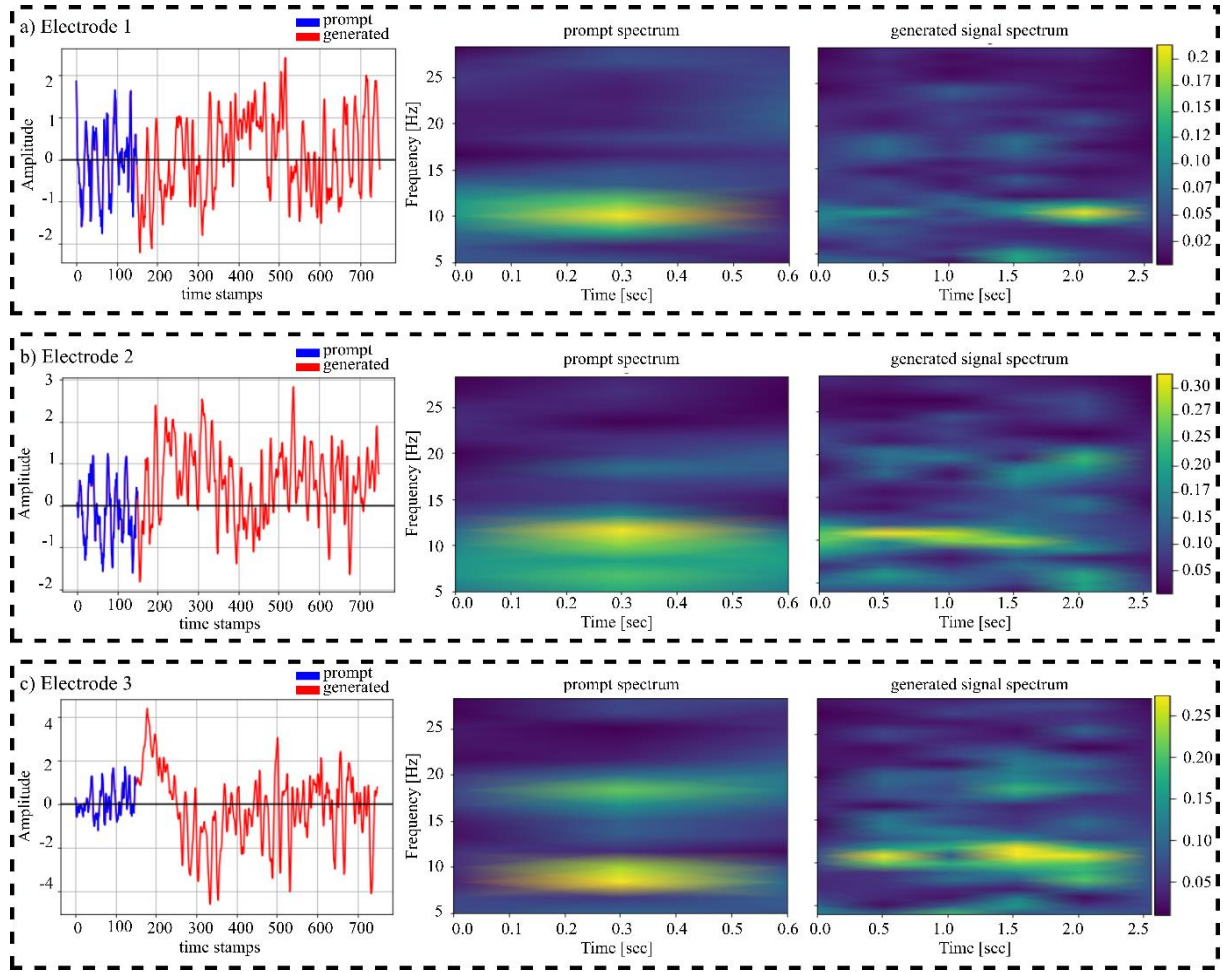

*Figure 12: Example 2. Multivariate signal generation of neural signals with unfiltered prompt of 150 samples. a), b) and c) show an example of the multi-electrode neural signal generation based on the unseen prompt from alpha-EEG dataset. Here each row corresponds to the neural signal generation of the respective electrode. For each electrode, the first column represents the input prompt in blue whereas the generated signal in red color. Columns two and three show the frequency spectra of input prompts and the generated signals respectively.*

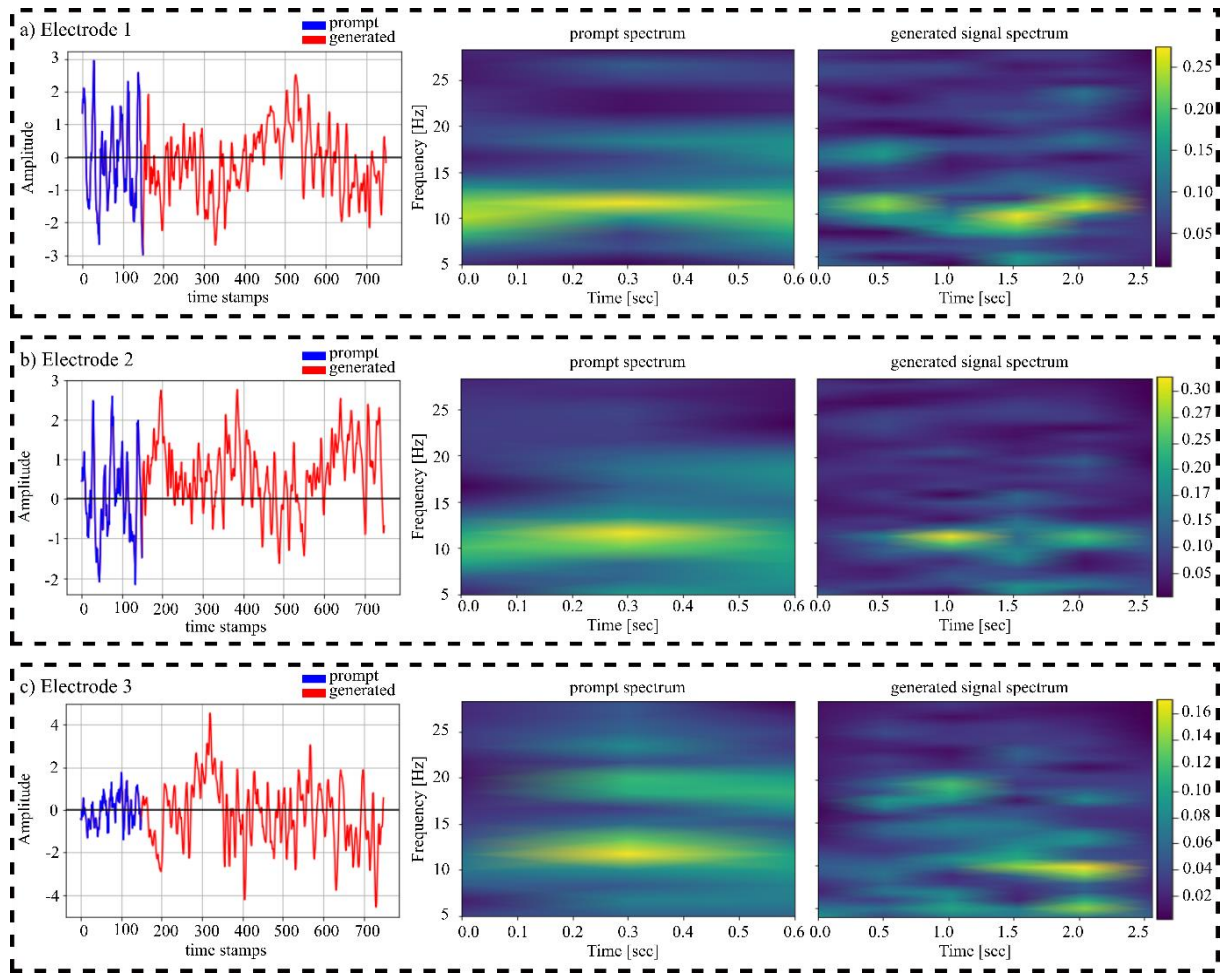

*Figure 13:* *Example 4. Multivariate signal generation of neural signals with unfiltered prompt of 150 samples. a), b) and c) show an example of the multi-electrode neural signal generation based on the unseen prompt from MI-EEG dataset. Here each row corresponds to the neural signal generation of the respective electrode. For each electrode, the first column represents the input prompt in blue whereas the generated signal in red color. Columns two and three show the frequency spectra of input prompts and the generated signals respectively.*

### 6.2.1 Comparison with ground truth and statistical analysis

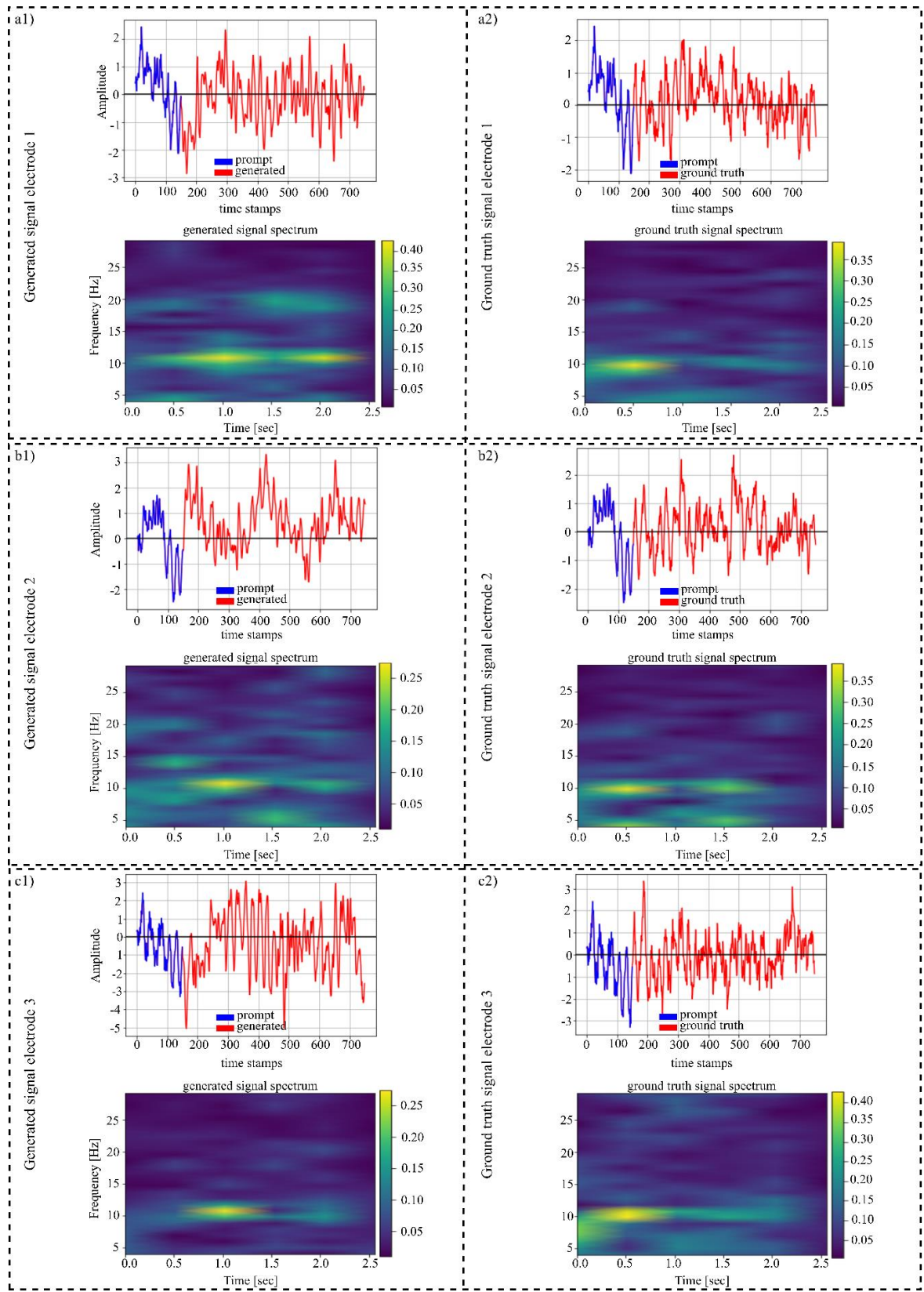

*Figure 14: Comparison between generated signals and the ground truth signals based on unseen prompts. a1, b1 and c1) represent the signals generated by GET and their corresponding spectra whereas a2, b2 and c2) represent the ground truth signals and their corresponding spectra.*

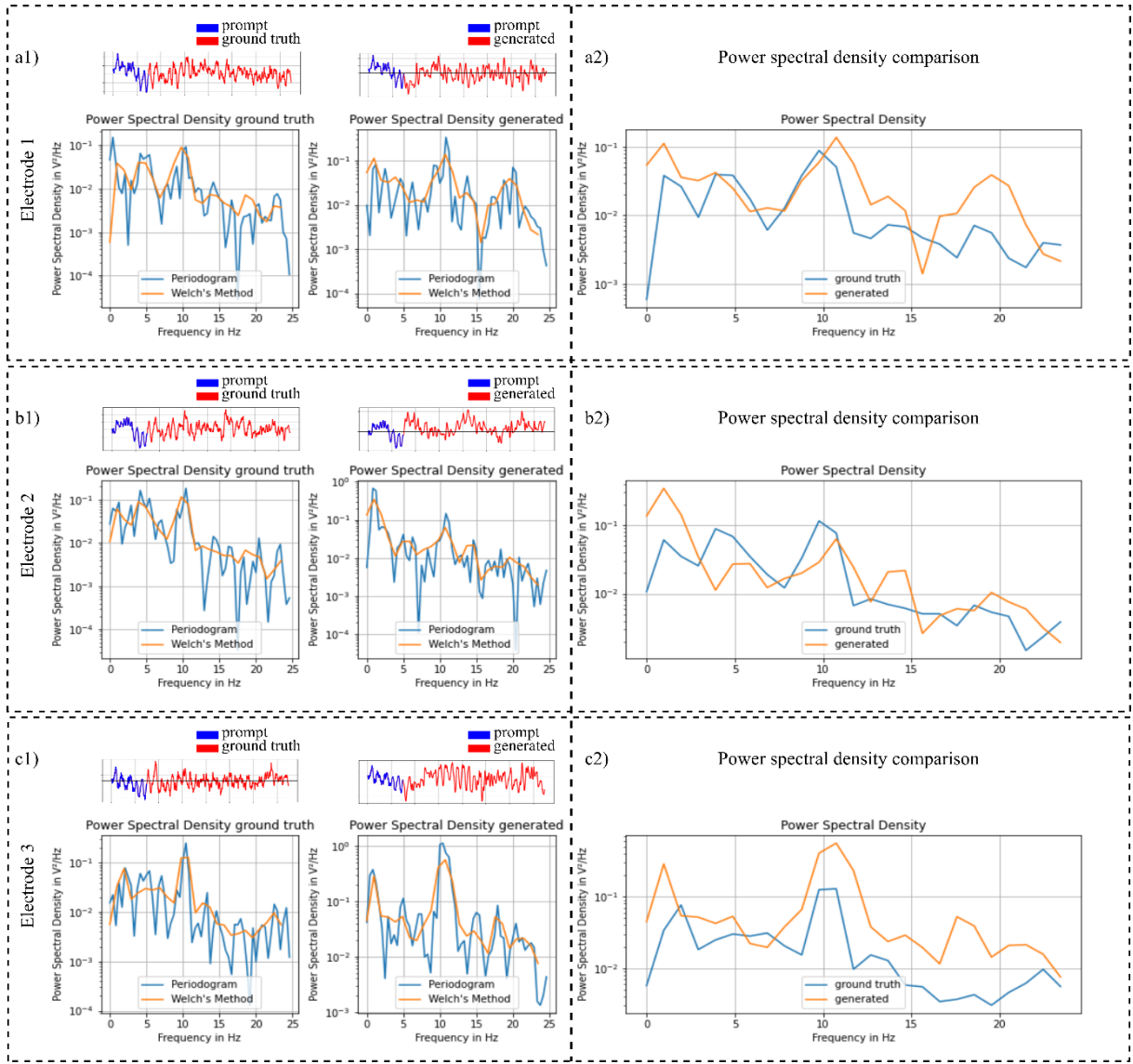

***Figure 15:*** *PSD of the generated signals and ground truth signals based on the given prompt. a1, b1 and c1) show the power spectral densities of the generated and ground truth signals computed using two different methods. a2, b2 and c2) show the comparison between the PSD of generated and ground truth signals.*

***Table 3:*** *Mean squared error (mse) of the PSD of ground truth and generated signal.*

| | mse | a2) 0.000914 | b2) 0.004983 | c2) 0.015759 |
|---|---|---|---|---|